\documentclass[aps,twocolumn,superscriptaddress,longbibliography]{revtex4-1}

\usepackage{amsmath}
\usepackage{amssymb}
\usepackage{amstext}
\usepackage{amsopn}
\usepackage{amsfonts}
\usepackage{amsxtra}
\usepackage[english]{babel}
\usepackage{graphicx}
\usepackage{bm}
\usepackage{multirow}
\usepackage{dcolumn}
\usepackage{color}
\usepackage{hyperref}
\usepackage{lineno}
\usepackage{todonotes}
\usepackage{verbatim}
\usepackage[acronym]{glossaries}
\usepackage{soul} % use \hl
\def\mathbi#1{\ensuremath{\textbf{\em #1}}}
\def\Q{\mathbi{Q}}
\def\qx{\ensuremath{q_{\parallel}}}
\def\qz{\ensuremath{q_{\perp}}}

\newacronym{LCCO}{LCCO}{La$_{2-x}$Ce$_{x}$CuO$_{4}$}
\newacronym{NCCO}{NCCO}{Nd$_{2-x}$Ce$_{x}$CuO$_{4}$}
\newacronym{RIXS}{RIXS}{resonant inelastic x-ray scattering}
\newacronym{ARPES}{ARPES}{angle-resolved photoemission spectroscopy}
\newacronym{STS}{STS}{scanning tunneling spectroscopy}
\newacronym{EELS}{EELS}{electron energy loss spectroscopy}
\newacronym{2D}{2D}{two-dimensional}
\newacronym{3D}{3D}{three-dimensional}
\newacronym{BZ}{BZ}{Brillouin zone}
\newacronym{FS}{FS}{Fermi surface}
\newacronym{combi}{combi}{doping-concentration-gradient}
\newacronym{FWHM}{FWHM}{full-width at half maximum}
\newacronym{RPA}{RPA}{random phase approximation}
\newacronym{LEG}{LEG}{layered electron gas}
\newacronym{SM}{SM}{Supplemental Material}
\newacronym{XAS}{XAS}{x-ray absorption spectroscopy}
\newacronym{r.l.u}{r.l.u}{reciprocal space units}
\newacronym{DMFT}{DMFT}{dynamical mean field theory}
\newacronym{DFT}{DFT}{density functional theory}
\newacronym{PLD}{PLD}{pulsed laser deposition}

\begin{document}

\title{Doping evolution of the charge excitations and electron correlations in electron-doped superconducting \acrlong*{LCCO}}

\author{J. Q. Lin}
\affiliation{School of Physical Science and Technology, ShanghaiTech University, Shanghai 201210, China}
\affiliation{Beijing National Laboratory for Condensed Matter Physics and Institute of Physics, Chinese Academy of Sciences, Beijing 100190, China}
\affiliation{University of Chinese Academy of Sciences, Beijing 100049, China}

\author{Jie Yuan}
\affiliation{Beijing National Laboratory for Condensed Matter Physics and Institute of Physics, Chinese Academy of Sciences, Beijing 100190, China}

\author{Kui Jin}
\affiliation{Beijing National Laboratory for Condensed Matter Physics and Institute of Physics, Chinese Academy of Sciences, Beijing 100190, China}
\affiliation{University of Chinese Academy of Sciences, Beijing 100049, China}

\author{Z. P. Yin}
\email{yinzhiping@bnu.edu.cn}
\affiliation{Department of Physics and Center for Advanced Quantum Studies, Beijing Normal University, Beijing 100875, China}

\author{Gang Li}
\affiliation{School of Physical Science and Technology, ShanghaiTech University, Shanghai 201210, China}

\author{Ke-Jin Zhou}
\affiliation{Diamond Light Source, Harwell Science and Innovation Campus, Didcot, Oxfordshire OX11 0DE, United Kingdom}

\author{Xingye Lu}
\affiliation{Department of Physics and Center for Advanced Quantum Studies, Beijing Normal University, Beijing 100875, China}
\affiliation{Photon Science Division, Swiss Light Source, Paul Scherrer Institute, CH-5232 Villigen PSI, Switzerland}

\author{M. Dantz}
\affiliation{Photon Science Division, Swiss Light Source, Paul Scherrer Institute, CH-5232 Villigen PSI, Switzerland}

\author{Thorsten Schmitt}
\affiliation{Photon Science Division, Swiss Light Source, Paul Scherrer Institute, CH-5232 Villigen PSI, Switzerland}

\author{H. Ding}
\affiliation{Beijing National Laboratory for Condensed Matter Physics and Institute of Physics, Chinese Academy of Sciences, Beijing 100190, China}
\affiliation{University of Chinese Academy of Sciences, Beijing 100049, China}

\author{Haizhong Guo}
\affiliation{School of Physical Engineering, Zhengzhou University, Zhengzhou 450001,
China}

\author{M. P. M. Dean}
\email{mdean@bnl.gov}
\affiliation{Department of Condensed Matter Physics and Materials Science, Brookhaven National Laboratory, Upton, New York 11973, USA}

\author{X. Liu}
\email{liuxr@shanghaitech.edu.cn} 
\affiliation{School of Physical Science and Technology, ShanghaiTech University, Shanghai 201210, China}

\date{\today}

\begin{abstract}
Electron correlations play a dominant role in the charge dynamics of the cuprates. We use \gls*{RIXS} to track the doping dependence of the collective charge excitations in electron doped \gls*{LCCO}. From the resonant energy dependence and the out-of-plane momentum dependence, the charge excitations are identified as \gls*{3D} plasmons, which reflect the nature of the electronic structure and Coulomb repulsion on both short and long lengthscales. With increasing electron doping, the plasmon excitations show monotonic hardening in energy, a consequence of the electron correlation effect on electron structure near the \gls*{FS}. Importantly, the plasmon excitations evolve from a broad feature into a well defined peak with much increased life time, revealing the evolution of the electrons from incoherent states to coherent quasi-particles near the \gls*{FS}. Such evolution marks the reduction of the short-range electronic correlation, and thus the softening of the Mottness of the system with increasing electron doping.
\end{abstract}

\maketitle

\section{Introduction\label{Introduction}}
The electronic behavior of metals is usually described using the standard Fermi liquid theory in terms of a single-particle spectral function of well-defined electron-like quasiparticles and a two-particle excitation spectrum dominated by long-lived collective charge excitations called plasmons \cite{Pines1999theory, Mahan2010}. It is, however, generally agreed that Fermi liquid theory breaks down in the high-temperature superconducting cuprates and the search for a fully satisfactory replacement theory remains one of the most studied problems in condensed matter physics \cite{Keimer2015quantum}. Empirically, techniques such as \gls*{ARPES}, \gls*{STS} and quantum oscillations have given us a detailed picture of the cuprates' electronic band structure \cite{Damascelli2003Angle}. Developing a comprehensive picture of the collective charge excitation in the cuprates has proved much more challenging. A key issue is limited $\vec{q}$-space access.  Optical techniques are intrinsically limited to $\vec{q}\approx 0$ \cite{Basov2005, Devereaux2007Inelastic}. Although \gls*{EELS} has no such restriction in principle, accessing $c$-axis dispersion remains difficult \cite{Fink2001Electronic, Roth2014, Mitrano2018, Vig2017Measurement}. \gls*{RIXS} has recently observed dispersive charge excitations in electron-doped cuprates opening new routes to characterize the charge dynamics of the high temperature superconductors \cite{Lee2014,Ishii2014, Dellea2017, Hepting2018}. Importantly, RIXS also provides bulk sensitivity. Many properties of such excitations have, however, not been investigated in detail, especially compared to extensive \gls*{RIXS} studies of the magnetic response \cite{Braicovich2010Magnetic, Tacon2011, Dean2012Spin,  Dean2013Persistence, Dean2015}. This leads to contradictory explanations of the origin of the observed charge excitations in the cuprates either as intraband particle-hole excitations \cite{Ishii2014, Ishii2017}, new modes arising from symmetry breaking near quantum critical point \cite{Lee2014} or collective plasmon excitations \cite{Greco2016,  Bejas2017, Greco2018, Eremin2018,  Hepting2018}. It is vital to resolve the nature of these excitations, and in particular how the electronic state evolves upon doping, which can deepen our understanding of the electron correlation, including both the short- and long-range Coulomb interactions, in the cuprate unconventional superconductors.

%%%%%%%%%%%%%%%%%%%%%%%%%%%%%%%%%%%%%%%%%%
% Figure 1
%
\begin{figure*}
\includegraphics[width = 1\textwidth]{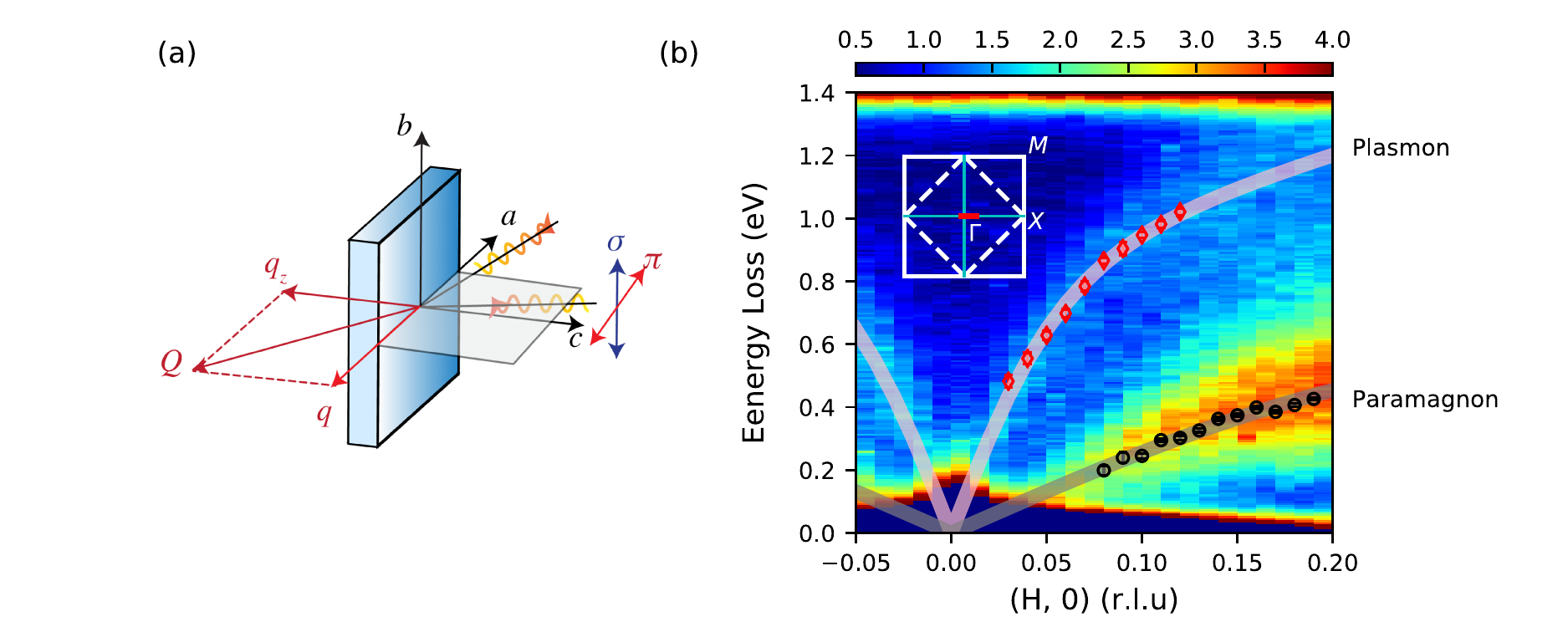}
\caption{Plasmon and paramagnon measurements of \gls*{LCCO} for $x=0.17$. (a) Sketch of experimental geometry. (b) \gls*{2D} color map of \gls*{RIXS} spectra with $2\theta = 146^{\circ}$. The elastic peak is subtracted for clarity. Red diamonds and black circles present the energies of the plasmon and paramagnon from fitting, respectively. The light pink line shows the fitting of the plasmon dispersion for a layered system (Eq.~\ref{eq:dispersion}) and the light black line arises from fitting a \gls*{2D} antiferromagnetic Heisenberg model to the paramagnon. The insert shows the two-dimensional \gls*{BZ} and the red line presents the measured range of in-plane wavevectors along the $(H, 0)$ direction.}
\label{Fig1}
\end{figure*}
%%%%%%%%%%%%%%%%%%%%%%%%%%%%%%%%%%%%%%%%%%
Here we report \gls*{RIXS} measurements on electron doped \gls*{LCCO} \cite{Jin2011} at the Cu $L_3$-edge, focusing on charge excitations around the two dimensional \gls*{BZ} center with various $c$-direction momentum transfer. By using \gls*{combi} films \cite{Yu2017}, we are able to provide much finer doping dependence than measured previously covering the optimal superconducting state to the non-superconducting metallic phase \cite{Lee2014, Ishii2014, Dellea2017,Hepting2018}. We find that the charge excitations have mostly collective nature, and exhibit out-of-plane momentum transfer dependence. This strongly supports the plasmon nature of the observed peaks, as does the fact that the dispersion of charge excitations can be empirically fitted by a \gls*{3D} plasmon model. As the electron doping increases from $x=0.11$ to $x=0.18$, the plasmon energy gently hardens and life time gradually increases. We explain these observations in connection with reduced effective mass and increased electronic quasiparticle coherence near the \gls*{FS}. Both are signs for a crossover from Mott physics to an itinerant picture, as doping drives the system from a high-temperature-superconducting to a metallic state. 

\section{Charge excitation and its plasmon nature oscillation \label{Plasmon Nature}}
Figure~\ref{Fig1}(b) plots Cu $L_3$-edge \gls*{RIXS} measurements of \gls*{LCCO} $x=0.17$. Distinct from measurements of hole-doped cuprates, where the low energy excitations are dominated by spin excitations \cite{Braicovich2010Magnetic, Tacon2011, Dean2012Spin,  Dean2013Persistence, Dean2015}, two low-energy dispersive modes are seen \cite{Lee2014, Ishii2014, Dellea2017, Hepting2018}. The excitation at lower energy with stronger intensity can be assigned as spin fluctuations, following previous studies of hole-doped cuprates \cite{Braicovich2010Magnetic, Tacon2011, Dean2012Spin,  Dean2013Persistence, Dean2015}. The other excitation with higher energy and weaker intensity was absent in  \gls*{RIXS} measurements of hole-doped systems, and appears to be unique to electron-doped cuprates \cite{Lee2014,Ishii2014, Dellea2017, Hepting2018}. This feature disperses much more steeply, and merges into the high energy $dd$ excitations above $\sim$ 1.3~eV with in-plane wavevector $\qx{} \geq 0.2$ \gls*{r.l.u}.     

\begin{figure}[ht]
\center
\includegraphics[width = 0.50\textwidth]{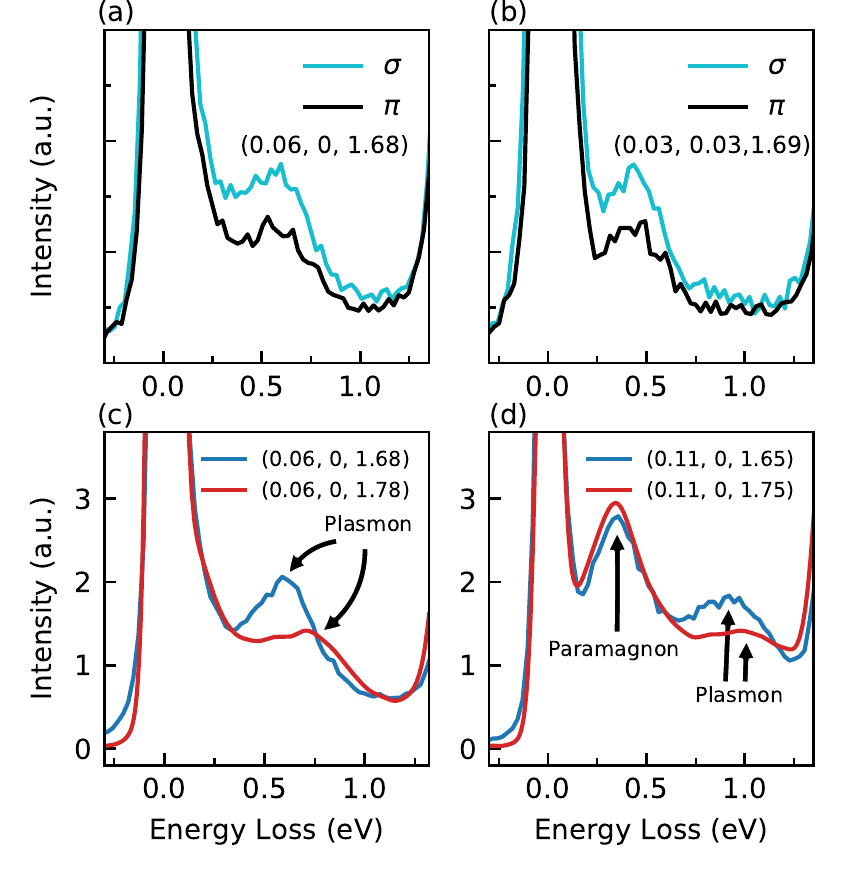}
\caption{Characterization of the plasmon nature of the charge excitation of \gls*{LCCO}. (a,b) Incident polarization dependent \gls*{RIXS} spectra at \qx{}=(0.06, 0) and (0.03, 0.03) for $x=0.18$. (c,d) Out-of-plane momentum transfer dependence of RIXS spectra at \qx{}=(0.06, 0) and (0.11, 0) for $x=0.17$. The paramagnon is independent of $L$ whereas the plasmon varies significantly with $L$.} 
\label{Fig2}
\end{figure}

To characterize this excitation, we first examined its polarization dependence. In the \gls*{RIXS} process, the x-ray photon can exchange angular momentum $m_J$ with the sample, thus excitations of different characters can be enhanced or suppressed in different polarization channels \cite{Luuk2009, Sala2011}. Figure~\ref{Fig2}(a) and (b) compare the excitation intensity with $\sigma$- and $\pi$-polarized incident beams at different \qx{} points. The strongly dispersive feature in Figure~\ref{Fig1}(b) consistently shows stronger spectral intensity with $\sigma$-polarized incident beam. Such polarization dependence indicates that the excitation is related to $\Delta m_J = 0$ or $2$ processes, namely pure charge excitations or bimagnon excitations. Bimagnon excitations, however, were suggested to disperse weakly near \gls*{BZ} center and appear as tails of the $\Delta m_J = 1$ paramagnon excitations \cite{Onose2004, chaix2018}. Therefore, the higher energy feature is most likely to be charge excitations, consistent with previous interpretations \cite{Lee2014,Ishii2014, Dellea2017, Hepting2018}.

%%%%%%%%%%%%%%%%%%%%%%%%%%%%%%%%%%%%%%%%%%
% Figure 2
%
%%%%%%%%%%%%%%%%%%%%%%%%%%%%%%%%%%%%%%%%%%

Further characterization of this excitation was provided by examining its out-of-plane wavevector \qz{} dependence, which probes the inter-CuO$_2$ plane coupling. In Figure~\ref{Fig2}(c) and (d), we compare spectra taken at different \qz{} with \qx{} fixed at different points. At $\qx = (0.11,  0)$, where both the spin excitation and the higher energy excitation can be clearly observed, the spin excitation energy stays constant for different \qz{}, consistent with the quasi-\gls*{2D} nature of this layered compound. On the other hand, the higher energy excitation behaves differently with strong \qz{} dependence \cite{Peng2017influence}. With a \qz{} variation from 1.68 to 1.78, its energy at $\qx = (0.06, 0)$ shifts by $\sim$~90~meV. Conversely, spin-related excitations are expected to show \gls*{2D} character. Thus the strong \qz{} dependence serves as a clear evidence supporting its charge fluctuation origin. Considering the strong dispersion of the charge excitations which extends $\sim$~1~eV in the observable \qx{} range, the natural choice of the leading interaction to host such excitations would be the long-range Coulomb interaction \cite{Hepting2018}. With a reasonable sized long-range Coulomb interaction to couple the electrons in different layers, plasmon excitations will become highly dispersive along the out-of-plane direction \cite{Sarma1982, Tselis1984, Jain1985, Sarma2009}.

\begin{figure}[h]
\center
\includegraphics[width = 0.48\textwidth]{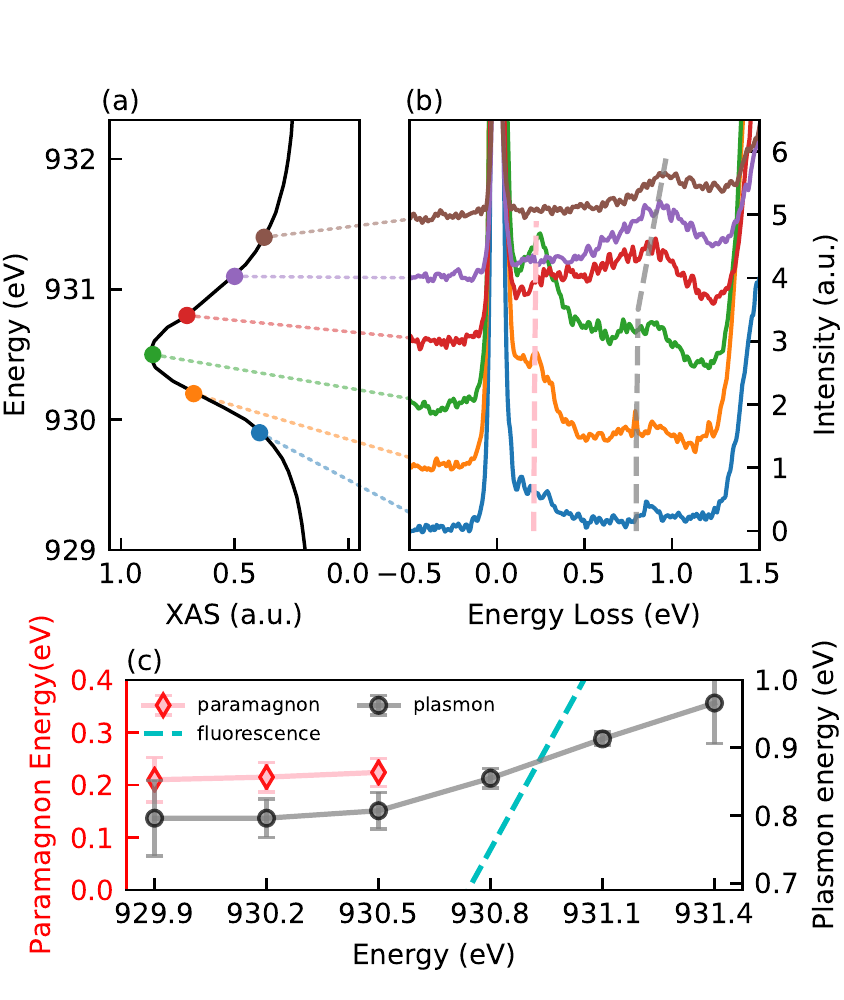}
\caption{Incident energy dependence of the plasmon implies a primarily collective nature of the excitation. (a) \gls*{XAS} data in total fluorescence yield mode for $x=0.17$. (b) \gls*{RIXS} spectra for $x=0.17$ at $\Q=(0.08, 0, 1.75)$ with different incident energies marked in (a), from $L_3-0.6$ to $L_3+0.9$~eV with step of 0.3~eV. Dashed lines indicate the paramagnon and plasmon energies. (c) Plasmon and paramagnon energies extracted from fitting. The dashed line indicates the expected incident energy dependence of constant final energy processes, such as fluorescence emission.}
\label{Fig3}
\end{figure}

To check the RIXS resonance behavior of the observed charge excitation, spectra at various incident x-ray energies across the Cu-$L_3$ edge were collected. Collective excitations are expected to exist at constant energy loss regardless of incident energy, whereas incoherent x-ray processes tend to occur at constant final energy \cite{Ament2011Resonant}.  As shown in Figure~\ref{Fig3}(b), both features from spin and charge excitations start to gain spectral weight at the rising edge of the Cu-$L_3$ edge white line, indicating they are stimulated by the same intermediate states, namely the ultra-fast transient double occupancy in the $2p^63d^9\xrightarrow{} 2p^53d^{10}\xrightarrow{} 2p^63d^{9\ast}$ RIXS process \cite{Dean2015}. The spin excitation spectral weight peaks at the maximum of the absorption curve, consistent with observations in hole-doped cuprates \cite{Minola2015}. Although both features share similar behaviors at the absorption rising edge, namely increasing in spectral weight and constant energy loss, they behave quite differently at the falling edge on the higher x-ray energy side. The spin excitation quickly diminishes, but the charge excitation seems to survive longer into the tail. More importantly, the charge excitation spectral weight reaches its maximum at $\sim$~0.3~eV above the energy of the white line peak and then gradually reduces, accompanied by a noticeable shift in its energy. We emphasize that, although the shift is obvious, the size of the shift is much smaller than the increase of the incident x-ray energy, so the data rule out constant final energy processes as shown in Figure~\ref{Fig3}(c). We assign the charge excitation as primarily a collective plasmon, but a small contribution of incoherent electron-hole processes appear to be present above the white line \cite{Minola2015,Minola2017, Huang2016raman}. It is worth noting that the observed \qz{} dependence is only expected within a plasmon model; electron-hole excitation would be expected to be independent of \qz{} \cite{Greco2016, Bejas2017, Greco2018}.

%%%%%%%%%%%%%%%%%%%%%%%%%%%%%%%%%%%%%%%%%%
% Figure 3
%

%%%%%%%%%%%%%%%%%%%%%%%%%%%%%%%%%%%%%%%%%%

\section{Plasmon doping dependence}
\begin{figure*}
\center
\includegraphics[width = 1\textwidth]{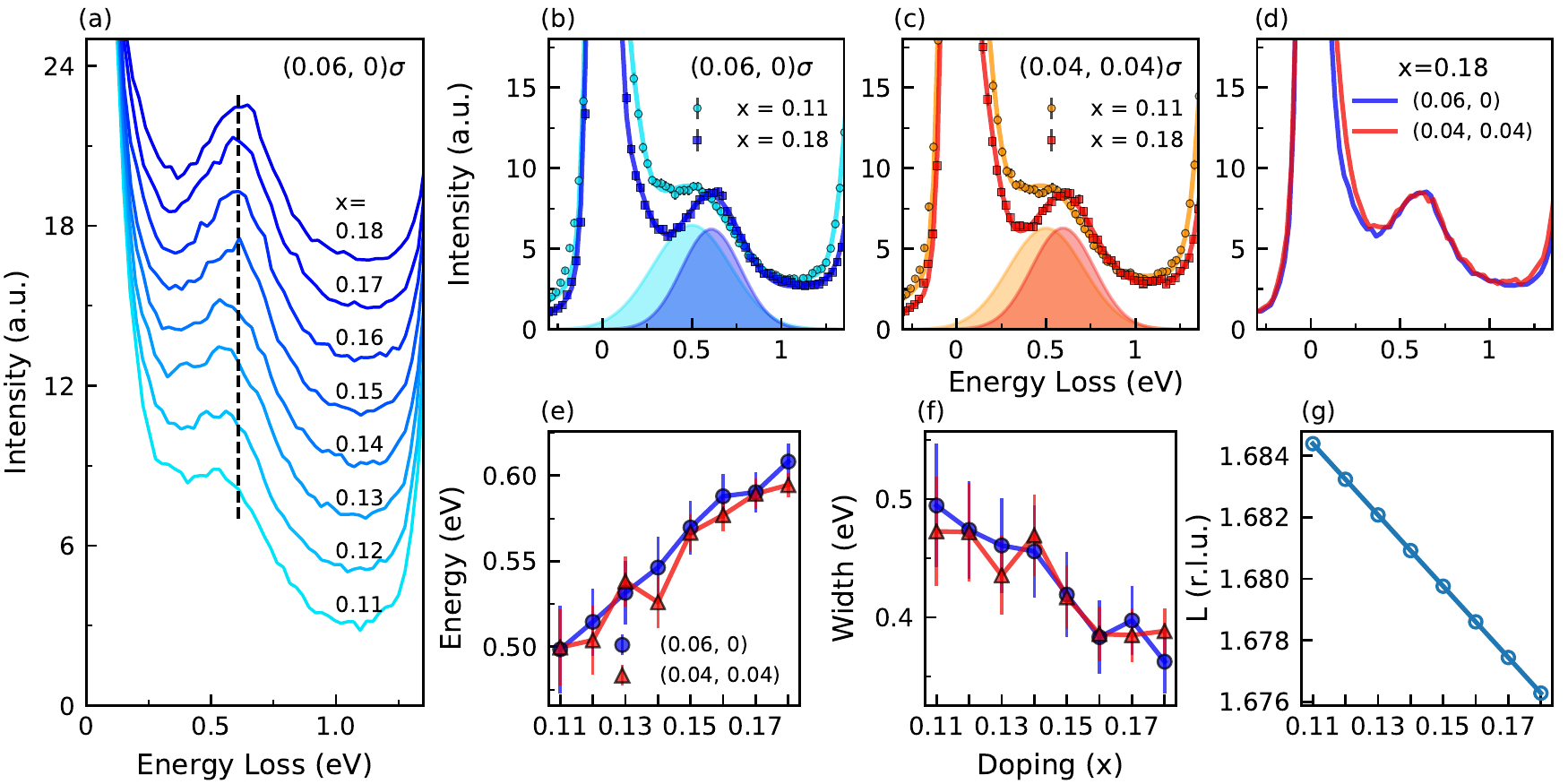}
\caption{Doping dependence of the plasmon. (a) \gls*{RIXS} spectra of \gls*{LCCO} at $\Q = (0.06, 0, 1.66)$ for $x=0.11$ to $0.18$. The vertical dashed line presents the plasmon energy for $x=0.18$ and spectra are offset vertically for clarity. (b,c) RIXS spectra at $\Q = (0.06, 0, 1.68)$ and $(0.04, 0.04, 1.68)$. In each panel, spectra for $x=0.11$ and $x=0.18$ are compared. The shaded peaks show the Gaussian profile representing the plasmon. (d) \gls*{RIXS} spectra measured at $\Q = (0.06, 0, 1.68)$ and $(0.04, 0.04, 1.68)$ show same dispersion with different in-plane directions $(H, 0)$ and $(H, H)$ for $x=0.18$. (e) Plasmon energy for each doping. (f) \gls*{FWHM}. (g) With fixed geometry, $L$ changes as a function of doping.}
\label{Fig4}
\end{figure*}

With the charge excitation identified as a plasmon, we investigate its doping dependence. By using the \gls*{LCCO} \gls*{combi} films, we were able to obtain a fine doping dependence by simply translating the sample along its doping gradient direction while keeping the experimental conditions identical. Four \qx{} points were studied, namely $(0.06, 0), (0.11, 0), (0.04, 0.04)$ and $(0.08, 0.08)$, which all show similar doping dependent behaviors. In Figure~\ref{Fig4}, data at $\qx = (0.06, 0)$ and $(0.04, 0.04)$ are shown as typical examples. More experimental results can be found in \gls*{SM} \cite{SM}.

%%%%%%%%%%%%%%%%%%%%%%%%%%%%%%%%%%%%%%%%%%
% Figure 4
%
%%%%%%%%%%%%%%%%%%%%%%%%%%%%%%%%%%%%%%%%%%

Figure~\ref{Fig4}(a) shows a stacking plot of the RIXS spectra at $\qx=(0.06, 0)$. Two direct observations can be made before any quantitative analysis. From $x=0.11$ to $0.18$, the plasmon excitation evolves from a broad feature to a well-defined peak, and its energy gradually hardens, consistent with studies of single phase samples \cite{Hepting2018}. Since the spectra were measured with identical experimental geometry, there is actually a small variation in the momentum transfer due to possible variation in lattice constants at different doping levels. As shown in Figure~\ref{Fig2}, the plasmon excitation energy strongly depends on \qz{} values. With this concern, the variation in the momentum transfer was carefully considered. Figure~\ref{Fig4}(g) shows the converted \qz{} values from the measured $c$-lattice constant variation of our combi film as a function of doping \cite{Yu2017}. From $x=0.11$ to $0.18$, the change in $c$-lattice constant is about 0.4$\%$, leading to a variation in \qz{} of 0.008. Importantly, Figure~\ref{Fig2} shows that, in the vicinity of these \qz{} values, the plasmon excitation hardens towards larger \qz{}. At higher doping level, towards which the measured RIXS signal hardens, the \qz{} actually reduces. Thus it is safe to conclude that the hardening upon further doping is an intrinsic property of the plasmon excitation. 

In Figure~\ref{Fig4}(b)-(c), \gls*{RIXS} spectra for two end doping levels, $x=0.11$ and $0.18$, are compared for clarity. The shaded peaks are the plasmon excitation components from our fitting \cite{SM}. The extracted energy and intrinsic, deconvolved peak width are shown in Figure~\ref{Fig4}(e)-(f). At both \qx{} points, the plasmon energy hardens by $\sim 20\%$ and the width sharpens by $\sim 20\%$. A variation in \qz{} of 0.1 at $\qx=(0.06, 0)$, as shown in Figure~\ref{Fig2}(c), leads to about 100 meV energy change in plasmon energy. A simple linear estimation suggests that the $c$-lattice constant reduction of 0.008 could lead to $\sim$~8~meV softening in the measured doping range, which is far smaller than the energy variation in Figure~\ref{Fig4}(e) and can be neglected compared to the much larger doping induced plasmon energy hardening.

We emphasize that the two \qx{} points we compare were specifically chosen such that their in-plane and out-of-plane momentum transfers are the same, although $(0.06, 0)$ is along the in-plane $(H, 0)$ direction and $(0.04, 0.04)$ is along the $(H, H)$ direction. The similarity of their RIXS response is further shown in Figure~\ref{Fig4}(d). At the same doping level, the charge excitation features for both \qx{} points almost overlap. Such $(H, 0) - (H, Hf)$ symmetry is confirmed at another $\vec{q}$ pair, namely $(0.11, 0, 1.65)$ and $(0.08, 0.08, 1.65)$. The connection of such observations to the band structure of \gls*{LCCO} will be discussed in section \ref{Sec:discussion}.

\section{Microscopic consideration of the collective excitations\label{Sec:Theory}}
A plasmon is an emergent collective mode in many-body physics arising due to the materials' electronic structure and Coulomb interactions including both the short-range interaction and the screened long-range interactions respectively \cite{Pines1999theory, Mahan2010}. The dispersion of the plasmon is given by the zero of dielectric function
\begin{equation}
\epsilon(\vec{q}, \omega) = 1 - v(\vec{q}) \Pi(\vec{q}, \omega),  
\label{eq1}
\end{equation}
where $v(\vec{q})$ is the Fourier transformation of Coulomb interaction, and $\Pi(\vec{q}, \omega)$ is the polarizability function.

Exact calculations of the plasmon dispersion in strongly correlated metals is challenging, partially due to the lack of knowledge on the vertex contribution which is needed to derive the polarizability function \cite{Mahan2010}. Progress has been made studying the 2D extended Hubbard model \cite{Loon2014} and the large-$N$ expansion of the layered $t-J-V$ model \cite{Greco2016}. In particular, the plasmon excitation energy was suggested to be closely related to the single-particle density distribution. The spectral weight and the renormalization of its dispersion reflect the evolution of the on-site Coulomb repulsion $U$ and the long-range screened Coulomb interaction $V$. Thus RIXS could serve as a good bulk sensitive tool to probe the electronic structures of the cuprates, complimentary to other spectroscopic techniques. 

Although the plasmon excitation in strongly correlated systems involves complicated processes, H. Hafermann \textit{et al.}\ \cite{Hafermann2014} pointed out that the polarizability $\Pi(\vec{q}, \omega)$ shares similar $\Pi(\vec{q}, \omega)\sim \alpha^2\frac{\vec{q}^2}{\omega^2}$ behavior as the Lindhard function in \gls*{RPA} calculations, in the long-wavelength limit. Such a property of the polarization function in the long-wavelength limit, even for the correlated system, is a consequence of gauge invariance and local charge conservation, regardless of the interaction strength \cite{Loon2014, Hafermann2014}. Thus for small \qx, to leading order, the electronic correlations only modify the scaling factor $\alpha$. This was numerically confirmed by the consistency of \gls*{RPA} and Dual-Boson calculations, where the vertex corrections from the \gls*{DMFT} local contribution was taken into account only in the latter \cite{Loon2014, Hafermann2014}. Applying this approach, we investigate the dispersion of the observed charge excitation with $\Pi(q, \omega)\sim \alpha^2 \frac{q^2}{\omega^2}$ in the low $\qx \lessapprox 0.13$ r.l.u regime that we focus on here, and treating the Coulomb potential as that from charge on a lattice \cite{Greco2016, Becca1996}. As a result the plasmon dispersion for a layered correlated system in the long wave-length limit can be approximated as
\begin{eqnarray}
\omega_{p}(\vec{q}) = \alpha A q_{\parallel}  \bigg [ & \frac{\epsilon_{\parallel}}{a^2}(2-\cos{aq_x}-\cos{aq_y}) + \nonumber \\* 
& \frac{\epsilon_{\perp}}{d^2}(1-\cos{dq_z}) \bigg ] ^{-1/2} . 
\label{eq:dispersion}
\end{eqnarray}
This includes material parameters $a$($d$) representing in-plane (inter-plane) lattice constant, $\epsilon_{\parallel}$($\epsilon_{\perp}$) the dielectric constants parallel (perpendicular) to the plane, and $A = \sqrt{\frac{e^2}{2a^2d}}$. Equation~\ref{eq:dispersion} was used to fit the measured plasmon dispersion with free parameters $p_1 = \alpha/\sqrt{\epsilon_{\parallel}}$
and $p_2 = \epsilon_{\parallel}/\epsilon_{\perp}$. The fitting for $x=0.17$ is plotted as the light pink line in Figure~\ref{Fig1}(c), which agrees with the RIXS measurements quite well. With the fitted $p_1$ and $p_2$, the plasmon energy for $x=0.17$ at $\vec{q}=0$ is calculated to be 1.24~eV, which is consistent with infrared optical and \gls*{EELS} measurements \cite{Onose2004, Nucker1989, Mitrano2018}. 

\begin{figure}
\center
\includegraphics[width = 0.48\textwidth]{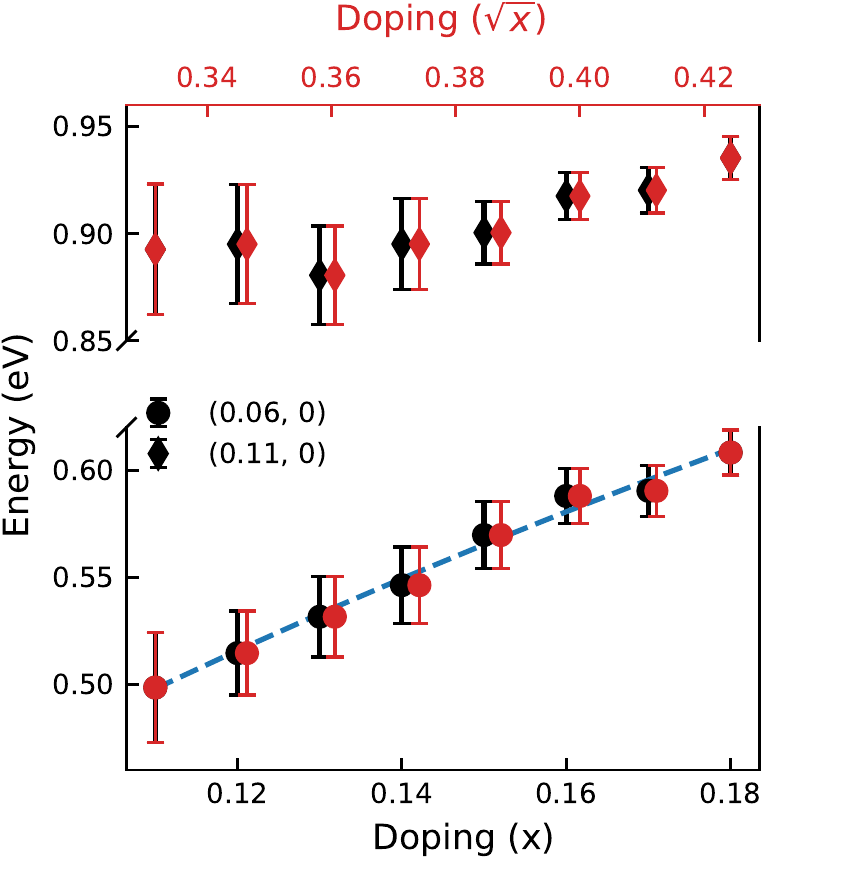}
\caption{Scaling of the plasmon energy versus $x$ and $\sqrt{x}$ for doping $x=0.11$ to $0.18$ at $\qx = (0.06, 0)$ and $(0.11, 0)$. The dashed blue line present a fit to $\omega_{p} = a \sqrt{x} + b$ with $a=1.19(7)$ and $b=0.11(3)$. Note that the non-zero $b$ value is inconsistent with simple fermi liquid model.}
\label{linearity}
\end{figure}

Within Eq.~\ref{eq:dispersion}, the doping dependence of the plasmon excitation is parameterized by  $\alpha$, which encapsulates the electron correlation effects. It is interesting to compare this coefficient with the naive expectation that the plasmon energy scales like the square root of the carrier concentration, as predicted from the free electron model. In Figure~\ref{linearity}, the extracted plasmon excitation energies are plotted as function of both $x$ and $\sqrt{x}$. At a lower \qx{} value of $(0.06, 0)$, the plasmon excitation could be described to be linear with $x$ or $\sqrt{x}$, with similar degree of agreement. Such ambiguity lies in the fact that the measured doping range is not large enough to clearly resolve any power law behavior (see appendix \ref{linearityF}). At a higher \qx{} value of $(0.11, 0)$, it is obvious that the plasmon excitation does not scale with either $x$ or $\sqrt{x}$. The disagreement with a simple free electron model comes as no surprise, and the degree of the disagreement serves as a measurement of the electron correlation effect in defining the electronic structures in the cuprates.    

% Figure 5

\section{Discussion\label{Sec:discussion}}
To help understand our observations, we examine the charge susceptibility via the Lindhard-Ehrenreich-Cohen expression for the charge dynamical susceptibility
\begin{equation}
    \chi_{0}(\Vec{q},\omega) \sim \sum_{\vec{k}}n_{\vec{k}}(1-n_{\vec{k} + \vec{q}})\frac{E_{\vec{k}-\vec{q}}}{\omega^2-E_{\vec{k}-\vec{q}}^2},
\label{eq3}
\end{equation}
where $n_{\vec{k}}$ is the Fermi-Dirac distribution function, and $E_{\vec{k}-\vec{q}}$ is the excitation energy for a $|\vec{k}> \xrightarrow{} |\vec{k}+\vec{q}>$ transition. This equation can model the elements of our results that are consistent with an effective quasiparticle picture, such as how Eq.~\ref{eq:dispersion} can provide a reasonable description of the data. At low $\vec{q}$, bounded by $n_{\vec{k}}(1-n_{\vec{k} + \vec{q}})$, electron-hole pairs involving electrons deeply under the FS do not contribute to $\chi_0(\Vec{q},\omega)$. Thus the low \qx{} plasmon we probe here is governed by the dynamics of the electrons near the \gls*{FS}, which are most relevant to the conducting behavior including superconductivity. In the low \qx{} limit, we can approximate $E_{\vec{k}-\vec{q}} \approx \vec{v} \cdot \vec{q}$ where $\vec{v} = \nabla_{\vec{k}}E_{\vec{k}-\vec{q}}$. The effective summation area in 2D reciprocal space, which contains electrons that contribute to the summation, is proportional to $\vec{q} \cdot \vec{k}_F$. Thus $\chi_0(\vec{q},\omega)$ is expected to be proportional to $\vec{v} \cdot \vec{k}_F$. It is worthwhile to clarify such \gls*{FS} sensitivity by considering the simple 2D electron gas case. Here the conducting band is defined as $E(\vec{k}) = \frac{k^2}{2m}$, $\vec{v} = \vec{k}_F/m$ and $k_F^2$ is proportional to total carrier density $n$. Thus, the low $\vec{q}$ plasmon intensity scales with $\frac{n}{m}$ consistent with general expectations. 

The above argument shows that the plasmon in \gls*{LCCO} measures the dot product of the Fermi velocity $\vec{v}$ with the Fermi wavevector $\vec{k}_F$, combined with the $\alpha$ factor (i.e.\ the prefactor in Eq.~\ref{eq:dispersion}) from the correlation effects. To appreciate the applicability of the quasiparticle description to the plasmon excitation and the doping dependent evolution of $\vec{v}$, the one particle spectrum for \gls*{LCCO} was calculated with \gls*{DMFT} combined with \gls*{DFT} \cite{Kotliar2006Electronic}. The main results are shown in Figure~\ref{DMFT}, and more details can be found in Supplemental Material \cite{SM}.
\begin{figure}
\center
\includegraphics[width = 0.48\textwidth]{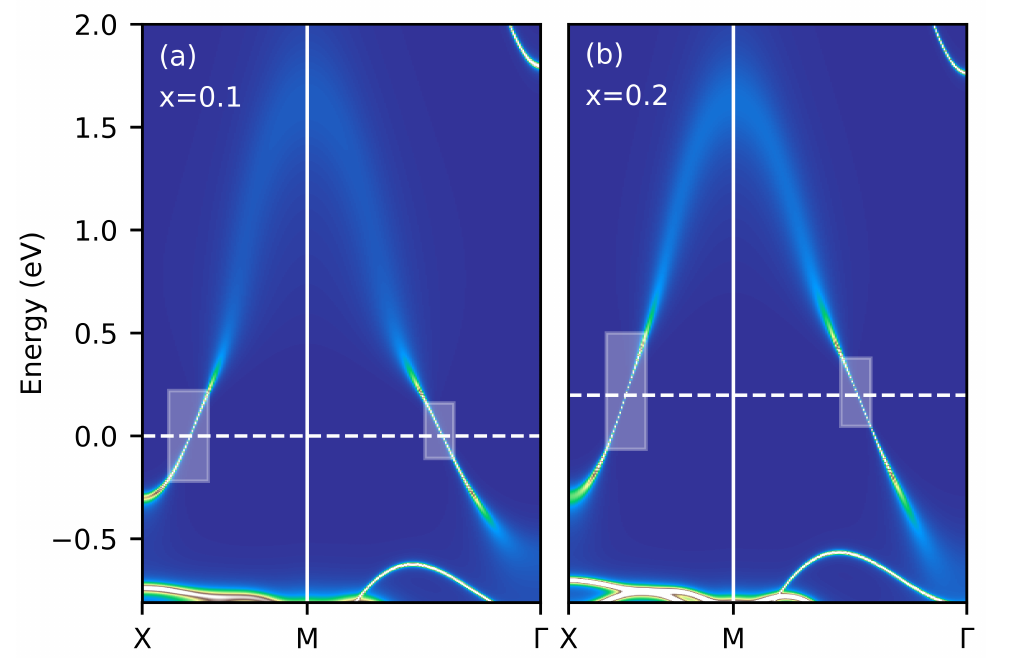}
\caption{Spectral function of \gls*{LCCO} from DMFT+DFT calculation for $x = 0.1$ (a) and $0.2$ (b). The square boxes highlight the electron and hole states that can be connected by momentum transfer $\qx = (0.06, 0)$.}
\label{DMFT}
\end{figure}

\begin{figure}
\center
\includegraphics[width = 0.48\textwidth]{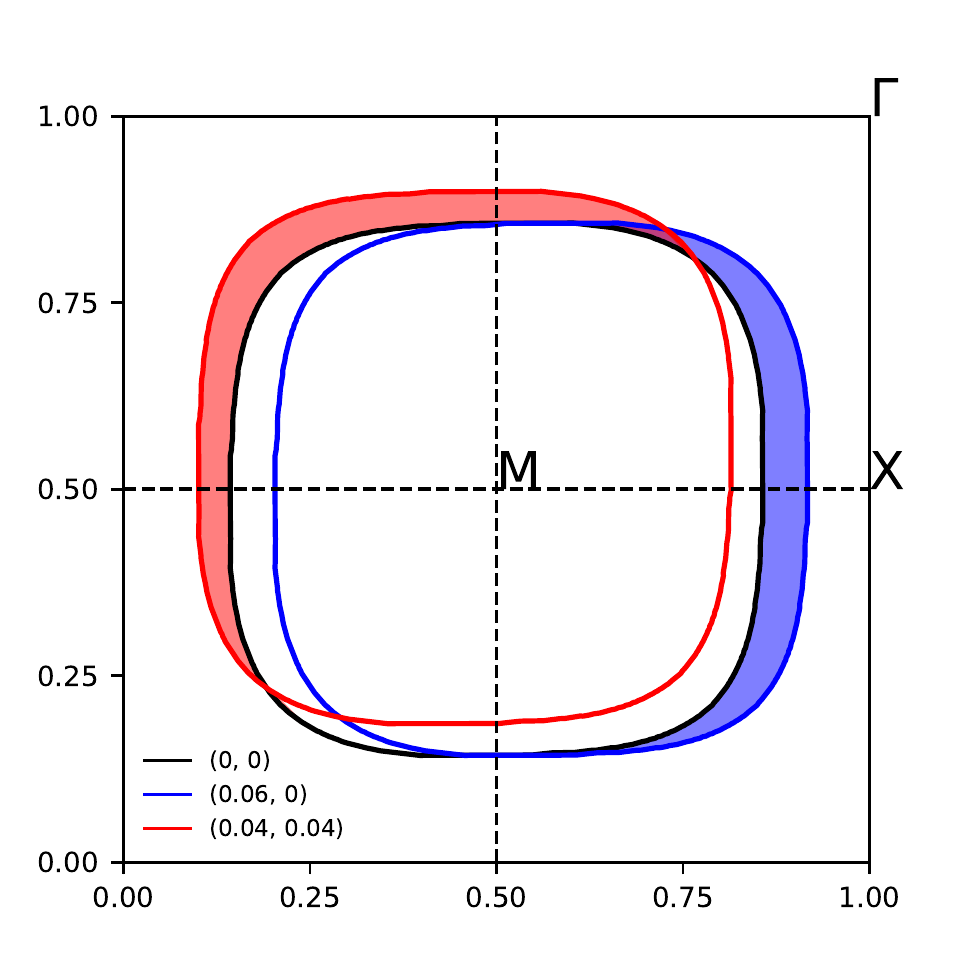}
\caption{Portions of the quasiparticle states that contribute to plasmon excitations at $\qx = (0.06, 0)$ and $\qx = (0.04, 0.04)$. The black circle shows the \gls*{FS} for $x = 0.1$ from DMFT+DFT calculations. The blue and red contours show \gls*{FS}s shifted by $\qx = (0.06, 0)$ and $\qx = (0.04, 0.04)$, respectively. The shaded blue and red regions show the electron states that can be excited into unoccupied hole states under momentum conservation constraints.}
\label{figure7}
\end{figure}
In the calculated spectrum, the electron correlation effects significantly reduce the quasiparticle lifetime away from the Fermi energy. While the states near the \gls*{FS} are of good coherent character. With increased electron itinerancy from $x = 0.1$ to $0.2$, the near-\gls*{FS} region that hosts long lived quasiparticles extends further out, as shown in Figure \ref{DMFT}. This explains the sharpening of the plasmon peak with increasing doping as observed in our \gls*{RIXS} data. As the plasmon excitation at low \qx{} originates from the dynamics of the electrons near the \gls*{FS}, its width measures the life time of the quasi-particles near the \gls*{FS} in the region accessible with $q_{\parallel}$. Upon increasing \qx{}, the plasmon excitation starts to probe the incoherent states that suffer more from the electron correlation effect (reduced lifetime and increased vertex corrections). Thus the excitation quickly broadens and become hard to identify in the \gls*{RIXS} spectrum as observed.

Figure \ref{figure7} gives a 2D view of the portion of the single particle spectrum where electrons can be excited into unoccupied hole states under momentum conservation constraints, as required in Eq.~\ref{eq3}. Although the plasmon is a collective excitation involving a large number of electrons, plasmon excitation along different $\vec{q}$ directions selectively probes different portion of the electron states near the \gls*{FS}. The almost overlapping plasmon excitations at $\qx = (0.06, 0)$ and $\qx = (0.04, 0.04)$, as shown in Figure~\ref{Fig4}, may suggest that the \gls*{FS} is more rounded than that from our DMFT+DFT calculations, consistent with \gls*{ARPES} reports \cite{Armitage2002Doping}. 

Following the above reasoning, we can speculate about the electron-hole asymmetry for the plasmon excitations which were, to the authors’ knowledge, never observed in Cu $L_3$-edge \gls*{RIXS} measurements of hole-doped cuprates, despite extensive studies of these materials \cite{Braicovich2010Magnetic, Tacon2011, Dean2012Spin,  Dean2013Persistence, Dean2013High, Dean2015, Minola2015, Minola2017}. The correlation effects could be still significant in the studied doping range, and coherent particles only live in a very small region in reciprocal space. Thus the plasmon excitation is expected to be weaker and broader, making it difficult to identify in the \gls*{RIXS} spectrum. Further, carriers in the hole-doped cuprates are known to have primarily oxygen rather than copper character, so O $K$-edge \gls*{RIXS} might be insightful \cite{Ishii2017}. Another consideration is the weaker intensity for charge excitations in hole-doped cuprates than electron-doped cuprates caused by the \gls*{RIXS} process and the asymmetric electronic states \cite{Tohyama2015}.

The hardening of the plasmon excitation upon increased doping is a combined effect from multiple factors. In the standard Ferimi liquid model, the plasmon excitation is expected to harden upon doping an electron pocket, as both $\vec{k}_F ^2$ and $\vec{v}$ are expected to increase. On the other hand, as discusssed in section \ref{Sec:Theory}, our observed hardening does not follow simple $\sqrt{x}$ relation, emphasizing the contribution from the local correlation effect. It would be interesting to compare the RIXS measured plasmon spectrum to that from the 2D extended Hubbard model \cite{Loon2014}. From the doping dependent evolution of the plasmon excitation lifetime and energy, the two major observations we made here, a full picture of crossing over from a strongly correlated Mott system to itinerant electron liquid could be obtained.

\section{Conclusions}
Our \gls*{RIXS} measurements of \gls*{combi} films reveal the collective plasmon nature of the charge excitations in \gls*{LCCO}. The cuprate electronic structure and Coulomb interactions determine the plasmon behavior, and both effects are expressed in different ways in different $\vec{q}$ regions. In our measured small $\vec{q}$ region, the plasmon dispersion can be described by single coherent particle model and is mainly governed by the itinerant character of the electrons and the long-range Coulomb interaction, while the correlation effect serves as a renormalization to determine the plasmon energy scale. By fitting the dispersion with the \gls*{2D} layered model, we extract effective parameters to describe the electron liquid that hosts high-temperature superconductivity. Our observed out-of-plane momentum dependence, demonstrates the \gls*{3D} nature of the Coulomb interactions, bringing us to consider cuprates as more than a \gls*{2D} CuO$_2$ plane. The merits of appreciable momentum range and being bulk sensitive distinguish \gls*{RIXS} from other conventional techniques in probing the two-particle charge dynamics in the cuprates. With improved energy resolution to probe even smaller regions close to the \gls*{FS}, future \gls*{RIXS} measurements could provide unprecedented bulk sensitive information on the evolution of the electronic structure in the electron doped cuprates. 

\begin{acknowledgements}
We thank Zi-Yang Meng and Jiemin Li for helpful discussions and support related to this project. Work at ShanghaiTech University was supported by the ShanghaiTech University startup fund and partially supported by MOST of China under Grant No.\ 2016YFA0401000. JQL was also supported by the international partnership program of Chinese Academy of Sciences under Grant No. 112111KYSB20170059. MPMD acknowledges the support from the U.S.\ Department of Energy, Office of Basic Energy Sciences, Early Career Award Program under Award No.\ 1047478. Work at Brookhaven National Laboratory was supported by the U.S. Department of Energy, Office of Science, Office of Basic Energy Sciences, under Contract No.\ DE-SC0012704. Work at PSI is supported by the Swiss National Science Foundation through the Sinergia network Mott Physics Beyond the Heisenberg Model (MPBH). Xingye Lu acknowledges financial support from the European Community’s Seventh Framework Programme (FP7/20072013) under Grant agreement No. 290605 (Cofund; PSI-Fellow). MD was partially funded by the Swiss National Science Foundation within the D-A-CH programme (SNSF Research Grant 200021L 141325). KJ acknowledges the financial support from CAS interdisciplinary innovation team and National Key Basis Reaserch Program of China grant 2016YFA0300301, 2017YFA0303003, 2017YFA0302902. ZPY was supported by the NSFC (Grant No.\ 11674030), the Fundamental Research Funds for the Central Universities (Grant No.\ 310421113), the National Key Research and Development Program of China grant 2016YFA0302300. The calculations used high performance computing clusters at BNU in Zhuhai and the National Supercomputer Center in Guangzhou. We acknowledge Diamond Light Source for time on Beamline I21 and the Swiss Light Source at the Paul Scherrer Institut for beamtime at ADRESS under Proposal 20151449.

\end{acknowledgements}

\appendix

\section{Sample\label{Sample_growth}}
A \gls*{LCCO} doping-concentration-gradient film was grown by \gls*{PLD} at the Institute of Physics, Chinese Academy of Sciences.  In our measured sample, $x$ varies from 0.10 to 0.19 along the chosen direction from edge to edge. Extensive characterization of this sample has been reported elsewhere \cite{Yu2017}. The high quality of the sample was also evidenced by the smooth evolution of the measured \gls*{XAS} spectrum during our experiments (see \gls*{SM} Figure~S1) \cite{SM}. The choice of such gradient film allows a careful doping dependent survey in fine doping steps. More importantly, changing the doping concentration was done by simply translating the film along the doping gradient direction while keeping the experimental condition identical. Thus experimental error is largely minimized, and data for different dopings can be compared with great confidence. To avoid possible non-intrinsic edge effects from the growth, our experiments were focused on the doping range of 0.11 to 0.18, covering the optimal superconducting state to the non-superconducting metallic phase \cite{Jin2011, Saadaoui2015, Yu2017}. The wave vectors used in our manuscript are indexed using the tetragonal ($I4/mmm$) space group with $a = b = 4.01$~\AA{}, $c$ continuously changes from 12.46~\AA{} for $x=0.10$ to 12.40~\AA{} for $x=0.19$.

\section{Experimental setup\label{Experimental_setup}}
Our experiments were carried out at the ADRESS beamline of the Swiss Light Source at the Paul Scherrer Institute \cite{strocov2010high, ghiringhelli2006saxes} and beamline I21 of the Diamond Light Source. The experimental geometry is sketched in Figure~\ref{Fig1}(a). Most of the data were taken with $\sigma$-polarized incident light, unless mentioned otherwise. At ADRESS, the scattered signal was collected at a fixed angle of 130 degrees, while this angle is of 146 degrees at beamline I21. This difference allows comparison of \gls*{RIXS} spectra at different out-of-plane momentum transfers (\qz{}) with the in-plane momentum transfer (\qx{}) fixed at certain points. At each station, the momentum transfer was varied by rocking the sample. Most of the \gls*{RIXS} spectra were collected with the incident x-ray photon energies at the maximum of the absorption curve near the Cu $L_3$-edge \cite{SM}, unless mentioned otherwise.

Along the gradient direction, the doping level $x$ varies from 0.10 to 0.19 in an 8~mm range, which is 0.011 per mm. The experiments were performed at the ADRESS beamline of the Swiss Light Source using the SAXES spectrometer and the I21 beamline of Diamond Light Source, with beamspots at the sample of $50(\text{H})\times 4(\text{V})$~$\mathrm{\mu m}^2$ and $30(\text{H})\times10(\text{V})$~$\mathrm{\mu m}^2$, respectively. Thus, the variation of doping level $x$ of the sample within the beamspot is less than 0.0006, after considering beam footprint effects and can be regarded as homogeneous. The instrumental resolution is about 120~meV at the ADRESS beamline and 50~meV at I21 beamline at the Cu $L_3$ edge, estimated from the \gls*{FWHM} of the elastic scattering from carbon tape. All data presented were collected at a temperature of 20~K. 

\section{plasmon dispersion fitting}
The total \gls*{RIXS} spectrum is fitted with five components: A pseudo-Voigt function for the elastic line, an anti-symmetrized Lorentzian function multiplied by Bose-Factor and convoluted with resolution function for the paramagnon, a Gaussian function convoluted with the resolution function for the plasmon and a Gaussian function for the $dd$ excitations. Background scattering is treated with a polynominal function. Similar approaches have been used extensively in the \gls*{RIXS} literature \cite{Tacon2011, Dean2013Persistence, Ishii2014, Hepting2018, Minola2015}. 

\section{linearity \label{linearityF}}
Assuming $\omega_{p}$ follows a linear relation to $x$ as,
\begin{equation}
    \omega_{p} = ax+b \xrightarrow[]{} \Delta\omega_{p} = a(x - x_0)
\end{equation}
Replace $x$ with ${\sqrt{x}}^2$, we have,
\begin{equation}
    \Delta\omega_{p} = a(\sqrt{x}-\sqrt{x_0})(\sqrt{x}+\sqrt{x_0})
\end{equation}
The variation of $(\sqrt{x}+\sqrt{x_0})$ in the measured range measures the error if a $\omega_{p}\sim \sqrt{x}$ relation is enforced. For $x$ from 0.11 to 0.18, $\sqrt{x}+\sqrt{x_0}$ is from 0.7125 to 0.8051 with $\pm 6\%$ variation from the averaged value, which is not big enough to be well resolved from the experimental data. 

\bibliography{reference}
\end{document}